\documentclass[aps,floats,twocolumn,epsf,prl,showpacs]{revtex4-1}
\usepackage{graphicx}
\usepackage{epstopdf}
\usepackage{hyperref}
\hypersetup{colorlinks=true,urlcolor=blue,citecolor=blue,linkcolor=blue,breaklinks=true}
\usepackage{color}
\usepackage[normalem]{ulem}

\begin{document}

\title{Generic optical excitations of correlated systems: $\pi$-tons}

\author{A. Kauch$^{a,*}$, P. Pudleiner$^{a,b,*}$, K. Astleithner$^{a}$, P. Thunstr\"om$^{c}$, T. Ribic$^a$, and K. Held$^a$}

\affiliation{$^a$Institute of Solid State Physics, TU Wien, 1040 Vienna, Austria}
\affiliation{$^b$Institute of Theoretical and Computational Physics,
Graz University of Technology, 8010 Graz, Austria}
\affiliation{$^c$Department of Physics and Astronomy, Materials Theory, Uppsala University, 751 20 Uppsala, Sweden}
\thanks{These authors contributed equally to this work}

\date{ \today }

\begin{abstract}
The interaction of light with solids gives rise to new bosonic quasiparticles, with the exciton being---undoubtedly---the  most famous of these polaritons. While excitons are the generic polaritons of semiconductors, we show that for strongly correlated systems another polariton is prevalent---originating from  the dominant antiferromagnetic or charge density wave fluctuations in these systems.
As these are usually associated with a wave vector $(\pi,\pi,\ldots)$ or close to it, we propose to call the derived polaritons $\pi$-tons.
These $\pi$-tons yield the leading vertex correction to the optical conductivity in all correlated models studied:
 the Hubbard, the extended Hubbard model, the Falicov-Kimball, and the Pariser-Parr-Pople model, both in the insulating and in the metallic phase.

\end{abstract}

\maketitle


\newcommand{\vek}[1]{\mathbf{#1}}
\newcommand{\Chi}{\protect\raisebox{2pt}{$\chi$}}
\newcommand{\kh}[1]{{\color{green}[KH: #1]}} 
\newcommand{\ak}[1]{{\color{blue}[AK: #1]}}
\newcommand{\pp}[1]{{\color{red}[pp: #1]}}
\newcommand{\akk}[1]{{\color{blue}{#1}}}

Since the springtime of modern physics, the interaction of solids with light has been of prime interest. The arguably simplest kind of interaction is Einstein's Noble prize winning photoelectric effect \cite{Einstein1905}, where the photon excites an electron across the band gap. 
More involved processes beyond a mere electron-hole excitation  can be described in general by effective bosonic quasiparticles, coined polaritons since  a polar excitation is needed to couple the  solid  to  light. 

\begin{figure}[t]
\includegraphics[width=0.51\linewidth,trim=5.8cm 9.5cm 5.8cm 8.5cm,clip]{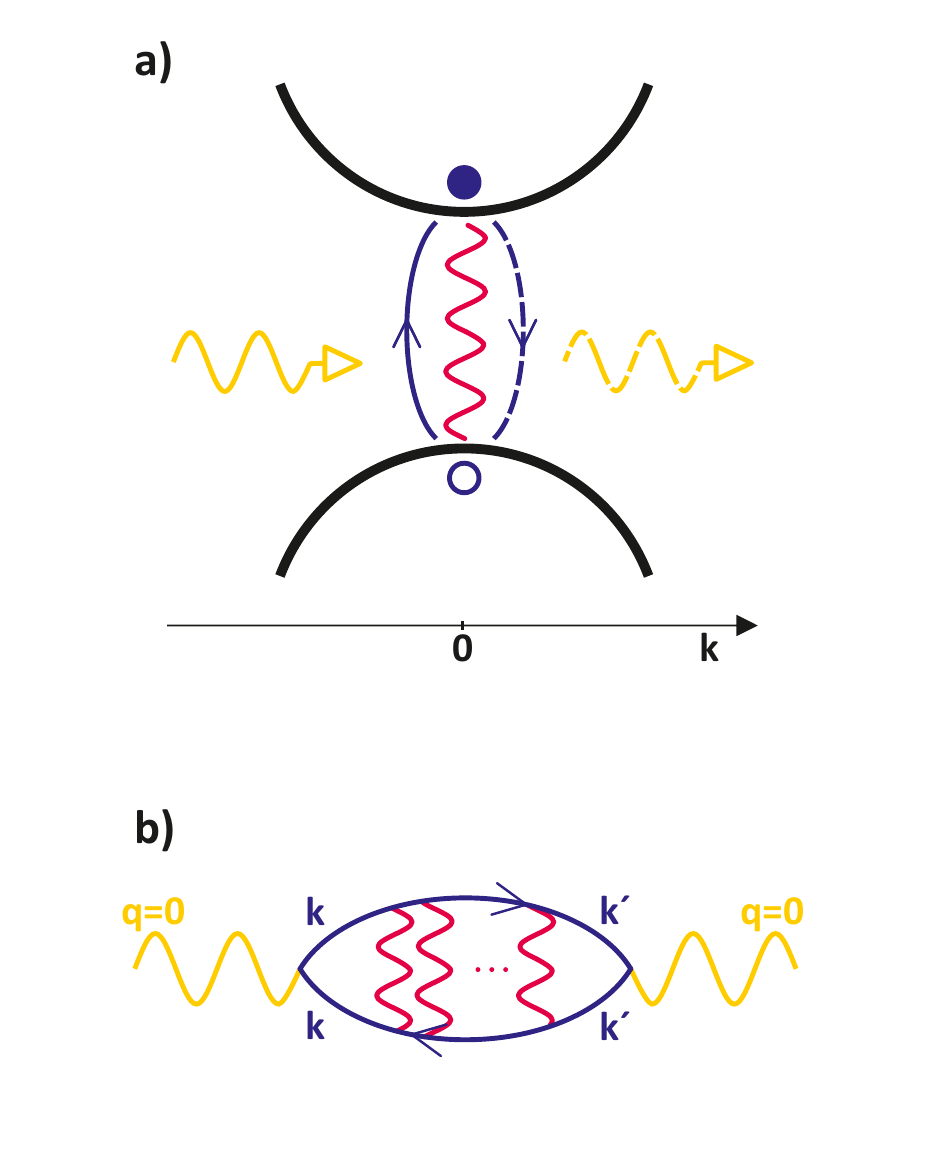}
\hspace{-.05\linewidth}  
\includegraphics[width=0.51\linewidth,trim=0.35cm 0cm 0.35cm 0.35cm, clip ]{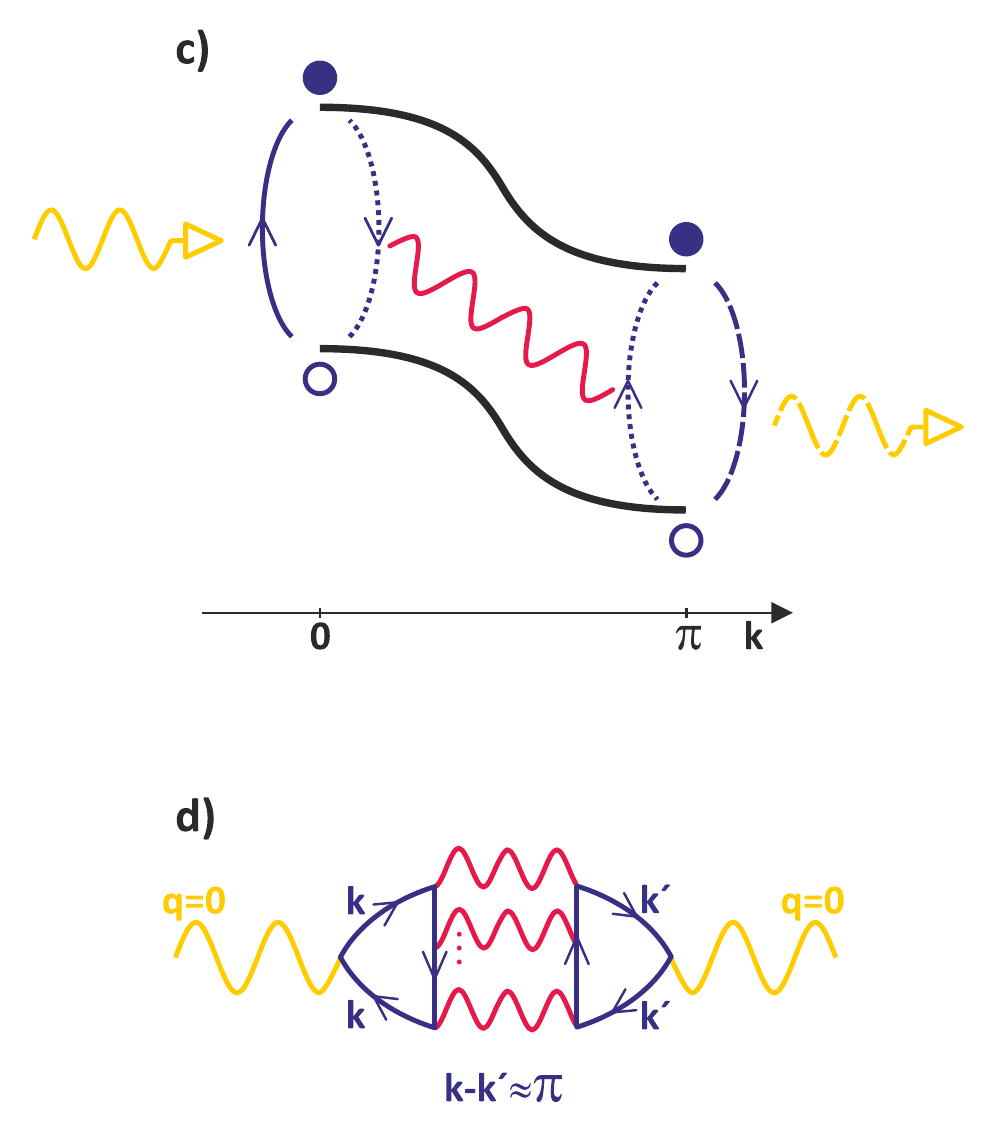}
   \caption{(Color online) Sketch of the physical processes (top) and Feynman diagrams (bottom) behind an exciton (left) and a $\pi$-ton (right). The yellow wiggled line symbolizes the incoming (and outgoing) photon which creates an
 electron-hole pair  denoted by open and filled circles, respectively. The Coulomb interaction between the particles is symbolized by a red  wiggled line; 
 dashed line indicates the recombination of the particle and hole; dotted line denotes the creation of a second particle-hole pair (right); black lines the underlying bandstructure (top panels).
 }
   \label{Fig:sketch}
\end{figure}

The prime example of a polariton is the exciton \cite{Frenkel1931,Wannier37}, where the excited electron-hole pair is  bound due to the Coulomb attraction between electron and hole. This  interaction is visualized in
Fig.~\ref{Fig:sketch} (a). 
Since it is an attractive interaction, an exciton requires the exciton binding energy less than  an unbound electron-hole pair. 
Other polaritons describe the coupling of the photon to surface plasmons, magnons or phonons.

 Fig.~\ref{Fig:sketch} (b) describes the exciton in terms of Feynman diagrams:  the incoming photon creates the electron-hole pair (distinguishable by the different [time] direction of the arrows) which  interact with each other repeatedly and finally  recombine emitting a photon. 
Since the energy-momentum relation of light is very steep compared to the electronic bandstructure of a solid, the transferred momentum from the photon is negligibly small ${\mathbf q}=0$. Thus,
electron and hole have the same momentum. For semiconductors this is often the preferable momentum transfer as well, connecting the bottom of the conductance with the top of the valence band as in  Fig.~\ref{Fig:sketch} (a).

In this paper we show that the generic polaritons for strongly correlated systems  are strikingly different. While semiconductors are band insulators with a filled valence and empty conduction band, strongly correlated systems are typically closer to a half-filled (or in general integer filled) band which is split into two Hubbard bands  by strong electronic correlations as visualized in  Fig.~\ref{Fig:sketch} (c) for a Mott insulator. (In case of a metallic system there is an additional quasiparticle band). Both metal and insulator are prone to strong  antiferromagnetic (AFM) or charge density wave (CDW) fluctuations with a wave vector close to  ${\mathbf q}=(\pi, \pi, \ldots)$ \cite{Keimer2013,Aeppli2014}. Indeed these
fluctuations can be described by the central part of the Feynman diagram 
 Fig.~\ref{Fig:sketch} (b), where the bare ladder diagrams correspond to the 
 random phase approximation (RPA).
However the  wave vector ${\mathbf q}=(\pi, \pi, \ldots)$  cannot directly couple to light, which only transfers  ${\mathbf q}=0$. Hence an exciton-like polariton as displayed in
 Fig.~\ref{Fig:sketch} (b) is not possible for AFM or CDW fluctuations.

As we will show in this paper, the  $(\pi, \pi, \ldots)$ fluctuations nonetheless constitute the dominant vertex corrections beyond a bare (bubble) particle-hole excitation. This is possible through a process where the central part of the Feynman diagram  Fig.~\ref{Fig:sketch} (b), i.e., the  $(\pi, \pi, \ldots)$  fluctuations,  are  rotated 
(and flipped) as sketched in Fig.~\ref{Fig:sketch} (d). Now it is possible to transfer   ${\mathbf q}=0$ from the photon and to pick up nonetheless the strong AFM or CDW fluctuations at  ${\mathbf k}-{\mathbf k}'\approx(\pi, \pi, \ldots)$. The physics of the associated process is visualized in   Fig.~\ref{Fig:sketch} (c). First, the light creates an electron-hole pair. Through the Coulomb interaction this electron hole-pair creates by impact ionization a second electron-hole pair  at a wave vector displaced by $(\pi, \pi, \ldots)$, and the two  interact repeatedly with each other, before emitting a photon again. Note that if one  assigns times to the electron-photon and Coulomb interactions in Fig.~\ref{Fig:sketch} (d) there are, after the first and till the last  Coulomb interaction, always two particle and two hole Green's functions [cf. Supplemental Material (SM) Fig. S8]. This  makes the $\pi$-ton distinctively different from Mott excitons \cite{Clarke1993,Wrobel2001,Essler2001,Jeckelmann2003} or quasiparticle-quasihole excitations, including those  envisaged in  \cite{Gull2009a} where the importance of AFM fluctuations was realized.

This excitation resembles to some extent~\footnote{Different is: in a superconductor  two particles interact through repeated ladders, here two electron-hole pairs  interact through a single ladder.} the pairing of electrons in superconductors through magnetic fluctuations. Since AFM or CDW fluctuations are typically at or close to a wave vector $(\pi, \pi, \ldots)$, we suggest to call this polariton a $\pi$-ton. But of course if a strongly correlated system happens to have its dominant fluctuations at another wave vector ${\mathbf k}-{\mathbf k}'\neq 0$, the same processes described in this paper allow for the coupling to light, creating polaritons.

In hindsight it appears rather obvious that AFM or CDW fluctuations couple this way to light. Why has this not been realized before? This is because  numerical methods such as quantum Monte Carlo~\cite{Gull2009a} or exact diagonalization~\cite{Mravlje2018} 
suffer from the difficulty to analyze the underlying physical processes highlighted by the parquet decomposition, and analytical methods as e.g. RPA or FLEX~\cite{Kontani2006} have been mostly biased with respect to certain channels such as the particle-hole ($ph$) channel in Fig.~\ref{Fig:sketch} (b) for excitons. Similar Feynman diagrams but with maximally crossed interaction lines, i.e., the particle-particle ($pp$) channel, have been made responsible for weak localization \cite{Altshuler1985} and strong localization \cite{Abrahams1979} in disordered systems. But the third (rotated) transversal particle-hole ($\overline{ph}$) channel of  Fig.~\ref{Fig:sketch} (d) has, to the best of our knowledge, not been considered hitherto, except for the second order diagram, the Aslamazov-Larkin  correction \cite{Aslamazov1968,Tremblay2011,Chubukov2014} which for half-filling compensates the second-order diagram of the $pp$-channel. Let us emphasize that it is however the whole ladder which is responsible for strong AFM or CDW fluctuations.

\begin{figure*}[tb]
\vspace{-0.2cm}
 \includegraphics[width=1.01\linewidth]{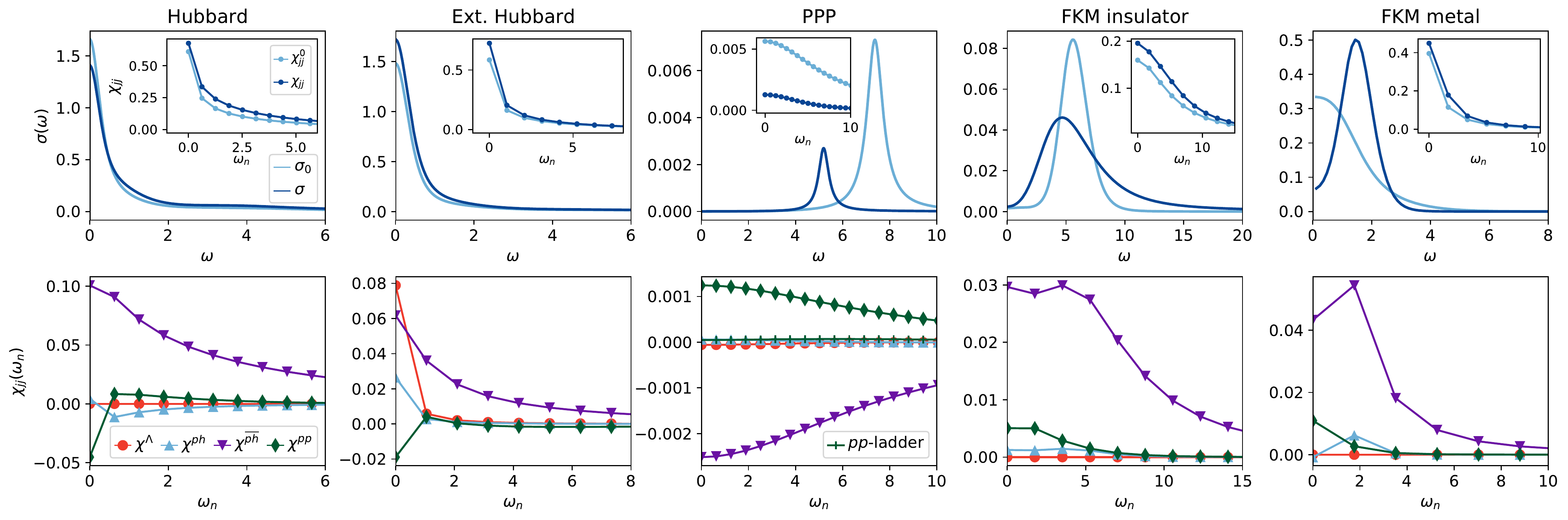}
   \caption{(Color online) Top: Optical conductivity for real frequency (main panel) and the corresponding current-current correlation function in Matsubara frequencies (insets) of the five cases studied, showing the bare bubble  ($\sigma_0$) and  the full conductivity ($\sigma$) including vertex corrections (in the insets $\Chi^0_{jj}$ and $\Chi_{jj}$, respectively). Bottom: Corresponding vertex correction to the current-current correlation function $\Chi_{jj}$
separated into  $ph$, $\overline{ph}$, $pp$ and $\Lambda$ contributions. For the PPP model also the contribution of a RPA-like $pp$-ladder is shown. 
Parameters from left to right: $U =4t$, $T=0.1t$ (HM); $U =4t$, $T=0.17t$,  $V=t$ (EHM); 
$T=0.1t$, $U =3.962t$, $V_{01}=2.832t$,   $V_{02}= 2.014t$,  $V_{03} = 1.803t$ (PPP; interactions translated into units of $t$ are 
fitted to experiment \cite{Bursill} ); $U=6t$, $T=0.28t$ (FKM insulator); $U=2t$, $T=0.28t$ (FKM metal).  
 }
   \label{fig:optcond}
\end{figure*}
Our insight has only been possible because of recent methodological advances which allow us to study all three aforementioned channels unbiasedly, using the parquet equations \cite{Bickers2004,Li2016,Li2017} within  the dynamical vertex approximation (D$\Gamma$A) \cite{Toschi2007,Kusunose2006,Katanin2009}, the dual fermion approach (DF) \cite{Rubtsov2008} and the parquet approximation (PA) \cite{Bickers2004}. For a review of these and related methods \cite{Rohringer2013,Taranto2014,Ayral2015,Li2015}, see \cite{RMPVertex}.

{\em Models and methods.} Let us now turn to the actual calculations, starting with introducing the models, which all can be summarized in the Hamiltonian
\begin{equation}
\! \! \mathcal{H} = -t \sum_{\langle i j\rangle\; \sigma}\!\! 
      c^{\dagger}_{i\sigma}c_{j\sigma}^{\phantom{\dagger}} + U\!  \sum_i n_{i\uparrow}n_{i\downarrow}
      + \frac{1}{2}\!\!\sum_{i\neq j, \sigma\sigma'}\!\!\!\! V_{ij}n_{i\sigma}n_{j\sigma'} \; 
    \label{eq:hamilton}
\end{equation} 
where $c^{(\dagger)}_{i\sigma}$ 
represents annihilation (creation) operator for an 
electron with spin $\sigma$ at site  $i$;  $n_{i\sigma}=c^{\dagger}_{i\sigma}c_{i\sigma}$; $\langle i j\rangle$ sums over each nearest neighbor pair $i,j$ once.  For the Hubbard model (HM) we have a local interaction $U$ only, i.e.,  $V_{ij}=0$, and $t$ denotes the hopping. We also study the extended Hubbard model (EHM), with nearest-neighbor interaction $V_{ij}=V$.  The Pariser-Parr-Pople model (PPP)~\cite{Pople53,Pariser53a} describes  conjugated $\pi$-bonds in carbon-based organic 
molecules and is here employed for a benzene ring, i.e., a one-dimensional chain with six sites, periodic boundary conditions and interactions between all sites.
Finally, the Falicov-Kimball model (FKM) \cite{Falicov1969,Freericks2003} has the same form as the  HM but the hopping is only for one spin species. All models  are solved for the square lattice (except PPP) at half-filling in the paramagnetic phase; $t\equiv 1$ and Planck constant $\hbar\equiv 1$ set our unit of energies and frequencies; for the optical conductivity lattice constant $a\equiv 1$, elementary charge $e\equiv 1$.

We employ the method which we consider most appropriate for the four models, i.e., the parquet D$\Gamma$A for the HM~\cite{Kauch2019}, the PA for the EHM and PPP (which is here more precise than a non-self-consistent D$\Gamma$A \cite{Pudleiner2019, Pudleiner2019a}), and a  parquet variant of the DF, extending earlier DF approaches~\cite{Otsuki2014,Jarrell2014, Ribic2016}.  We solve the parquet equations on a $6\times 6$ momentum grid, except for  the PPP for benzene which has 6 sites or momenta.  For the HM, EHM and PPP we use the {\em victory} code~\cite{Li2017} to solve the parquet equations, and {\em w2dynamics}~\cite{w2dynamics2018} to calculate the fully irreducible vertex in case of the HM; for the FKM we employ a reduced frequency structure of the vertex \cite{Ribic2016,Ribic2016b} implemented in a special-purpose parquet code~\cite{Astleithner2019}. 
~>
The optical conductivity 
   \mbox{$\sigma(\omega) 
     \! =\! \Re\,\frac{\Chi_{jj}^{\vek{q}=0}(\omega+i\delta)
                  -\Chi_{jj}^{\vek{q}=0}(i\delta)}{i(\omega+i\delta)}$},
 for $\delta\!\to\! 0$, is calculated from the current-current correlation function $\Chi_{jj}^{\vek{q}=0}$  at Matsubara frequency $\omega_n$ and momentum $\vek{q}=0$ , which can be separated into a bubble term consisting of two Green's functions $G_{k}$  only  and vertex corrections  $F^{kk'q}_{d}$ in the following way:
 \begin{eqnarray}
      \Chi_{jj,q} 
       &=& \frac{2}{\beta N}\sum_{k}
          \left[\gamma_{\vek{k}}^{\vek{q}}\right]^2
          G_{q+k}G_{k}  \nonumber \\
       &&+ \frac{2}{(\beta N)^2}\sum_{k,k'}
          \gamma_{\vek{k}}^{\vek{q}} \gamma_{{\vek{k}}'}^{\vek{q}}
          G_{k'}G_{q+k}
          F^{kk'q}_{d}
          G_{q+k'}G_{k} \;.
      \label{eq:ccCkw}
  \end{eqnarray}
Here, we use a four-vector notation $k=(\vek{k},\nu_n)$
with  $q=(\vek{q}=0,\omega_n)$; $\gamma_{\vek{k}}^{\vek{q}=0}=\partial\epsilon_{\vek{k}}/\partial \vek{k}$ denotes the dipole matrix elements given by the derivative of the energy-momentum relation in the Peierls approximation~\footnote{except for the  PPP model  where $\gamma_{\vek{k}}\equiv 1$, which corresponds to the dynamical compressibility; for additional results with $\gamma_{\vek{k}}^{\vek{q}=0}=\partial\epsilon_{\vek{k}}/\partial \vek{k}$ see SM.}; $\beta=1/T$ is the inverse temperature and $N$ the number of ${\mathbf k}$-points.

In the parquet-based approaches employed, the vertex $F$ contains contributions from the  fully irreducible vertex $\Lambda$ as well as contributions that are reducible in the  three channels ($ph$, $\overline{ph}$, $pp$): $F=\Lambda + \Phi_{ph} +\Phi_{\overline{ph}}+ \Phi_{pp}$  \footnote{For a similar analysis for the self-energy cf.~\cite{Gunnarsson2016}}. The density component $F_d$ that enters the optical conductivity denotes the even spin combination~\cite{RMPVertex,Li2016}.

 Inserting in Eq.~\ref{eq:ccCkw} instead of $F$ one of the summands $\Lambda$,  $\Phi_{ph/{\overline{ph}}/{pp}}$ we obtain the contributions from the respective channels: $\Chi^{\Lambda}$, $\Chi^{ph}$, $\Chi^{\overline{ph}}$, and $\Chi^{pp}$. 
 The most simple contributions to $\Chi^{ph}$ and  $\Chi^{\overline{ph}}$ are just the ladder diagrams  of Fig.~\ref{Fig:sketch} (b) and (d), respectively.
For the analytic continuation of the optical conductivity to real frequencies we employ the maximum entropy method~\cite{Kaufmann2018}; for the PPP we use Pad\'e interpolation. 

{\em Results: Optical Conductivity.} Let us now turn to the results, starting with the optical conductivity in Fig.~\ref{fig:optcond} (top). Within the four models, we studied five physically different examples: HM (metal), EHM (metal), PPP (insulator), FKM (insulator), and FKM (metal) [see SM for results at different parameters]. In all five cases we see noticeable vertex corrections. For the two insulators, especially for the PPP, there is a strong reduction of the optical gap compared to the one-particle gap (bare bubble contribution $\sigma_0$). Usually one would associate such a reduction to the exciton binding energy. However, when inspecting the contribution of the individual  channels  in Fig.~\ref{fig:optcond} (bottom),  we see that it is not the $ph$-channel of the exciton but the  $\overline{ph}$-channel which is  dominating and  responsible for the reduction of the optical gap. 
Note that a $ph$-ladder built from a local interaction (RPA) or a local vertex (as  e.g. in dynamical mean field theory \cite{Georges1996}) has zero contribution to the optical conductivity~\footnote{Because $G_{q+k}G_{k}$ in Eq.~(\ref{eq:ccCkw}) is even in $\vek{k}$ for ${\vek{q}=0}$ and $\gamma_{\vek{k}}^{\vek{q}=0}$ is odd}. This is why we included in our study also the PPP and EHM where through non-local interaction one obtains simple ladder contributions in the $ph$-channel~\footnote{The non-zero $ph$ contributions in the HM and FKM stem from insertions of other diagrams into the $ph$-ladder through the parquet equations.}.

For two of the metallic cases (HM and FKM) the vertex corrections reduce the conductivity at small frequencies. One might be tempted to associate this with weak localization corrections, i.e., the $pp$-channel. But again by inspecting the vertex contributions in Fig.~\ref{fig:optcond} (bottom) we see that it is the  $\overline{ph}$-channel that is dominating; the  $pp$ contribution is small. The third metallic case (EHM) is different in the sense that, besides the  $\overline{ph}$-channel, the bare vertex $\Lambda$ contributes to a similar amount. This is because the non-local interaction provides an additional way to polarize the system and hence to couple to light.

\begin{figure}
\vspace{-0.2cm}
 \includegraphics[width=\linewidth]{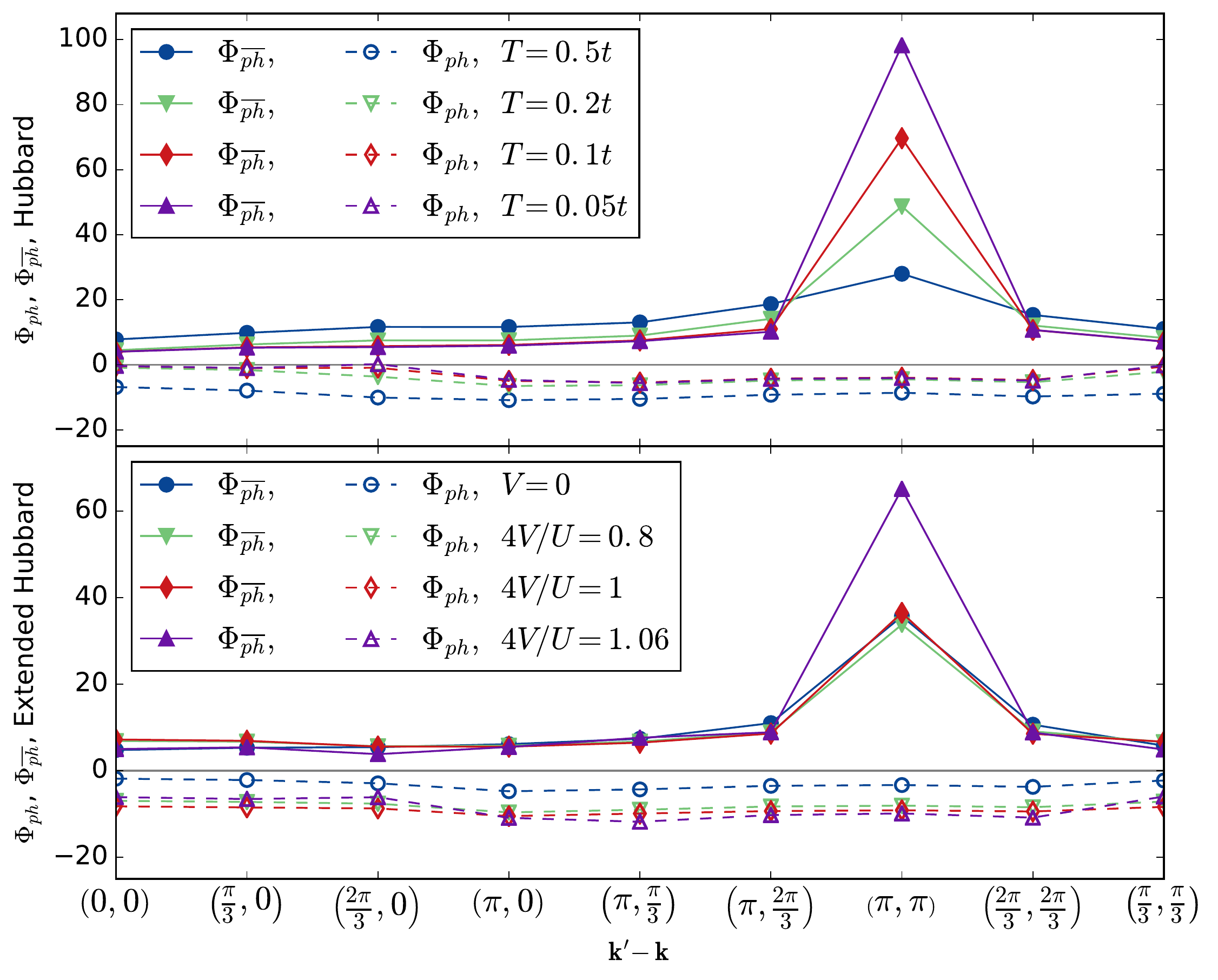}
   \caption{(Color online) 
Reducible contributions $\Phi$  in the $ph$ and $\overline{ph}$-channel to the full vertex $F_{d,{\vek{k}\vek{k}'\vek{q}=0}}^{\nu_n\nu_n'\omega_n}$ correcting the optical conductivity. Top: HM at various temperatures and $U=4t$. Bottom: EHM  at $U=4t$, $T=0.17t$ and various $V$. Shown is the contribution $\nu_n= \nu_n'= \pi T$; $\omega_n=0$ at fixed $\vek{k}=0$ as a function of $\vek{k}'-\vek{k}$.
   \label{Fig:phi_k}}
\end{figure}

In all cases except for the EHM, the  $pp$-channel provides the second largest  contribution. One might suspect that this stems from simple RPA-like ladder diagrams as envisaged in the theory of weak localization. But this is not the case. In the case where this  $pp$-channel is largest, i.e., for the PPP, we additionally plot the contribution from a bare RPA-like $pp$-ladder. It is negligibly small.


{\em Physical origin of vertex corrections.}
Why does the  $\overline{ph}$-channel give such a big contribution? It is because of the dominant fluctuations in the system. These are AFM or CDW fluctuations at a wave vector  $(\pi,\pi,\ldots)$ (see below). These fluctuations are already generated by RPA-ladder diagrams in the $ph$-channel and in  the $\overline{ph}$-channel as visualized in Fig.~\ref{Fig:sketch}. Let us emphasize however, that the employed parquet methods take many more Feynman diagrams and the mutual coupling of these channels into account. This coupling, in particular the  $pp$-inclusions, leads to a damping of the contribution of $ph$- and $\overline{ph}$-channel. One can still envisage the physics as in   Fig.~\ref{Fig:sketch} (d) but  with a renormalized  (screened) interaction  and a renormalized propagator. 

The fact that, on the other hand, the bare  $pp$-ladder in Fig.~\ref{fig:optcond} (bottom middle)  is small shows us that there is a strong feedback of the AFM or CDW fluctuations into the  $pp$-channel 
through the parquet equations, which leads to the considerable contributions of the $pp$-channel. In other words, these  $pp$ contributions arise (only) as a consequence of the enhanced AFM and CDW fluctuations.\footnote{In contrast, the 
 bare $ph$-ladder would be much larger without feedback from the other channels.}

To demonstrate the importance of the  $(\pi,\pi)$  contribution, we plot in  Fig.~\ref{Fig:phi_k}
the reducible contributions $\Phi$ to the full vertex $F$ as a function of $\vek{k}'-\vek{k}$  for the $ph$ and $\overline{ph}$-channel, setting $\vek{q}=0$ for the optical conductivity. Note that although the reducible $ph$ and $\overline{ph}$ vertices are interrelated,
it is a different momentum and frequency combination that enters the optical conductivity [see SM].
As we see in  Fig.~\ref{Fig:phi_k} this $ph$-contribution is small and the $\overline{ph}$-contribution is strongly peaked at the wave vector $\vek{k}'-\vek{k}=(\pi,\pi)$ because of the strong AFM and CDW fluctuations for the HM and EHM, respectively. A similar finding holds for the FKM and PPP model [see SM].
Hence we can conclude that it is indeed predominately the $\vek{k}'-\vek{k}=(\pi,\pi)$ contribution that is responsible for the vertex corrections in the optical conductivity, and therefore we call these polaritons $\pi$-tons.


{\em Characteristics of the $\pi$-ton.}
While AFM and CDW fluctuations are dominant at all parameters and temperatures analyzed, they become---as a matter of course---stronger when we approach a corresponding phase transition. This effect can be seen in Fig.~\ref{Fig:phi_k}
For the HM (top panel of  Fig.~\ref{Fig:phi_k}), reducing the temperature means that AFM fluctuations become strongly enhanced, cf. \cite{Katanin2009,Rohringer2011,Otsuki2014,Schaefer2016,Gukelberger2016}. While there is no finite-temperature phase transition in two dimensions, the correlation length becomes exponentially large \cite{Schaefer2015-2}.
For the EHM, Fig.~\ref{Fig:phi_k} (bottom), we instead enhance the non-local interaction $V$. This way we approach a phase transition towards CDW ordering (at $4V=U$ in the atomic limit and at a slightly larger $V$'s here~\cite{Pudleiner2019a}).

{\em Conclusion and outlook.} We have provided compelling evidence for what appears to be the generic polaritons in strongly correlated electron systems---at least in one and two dimensions. These polaritons, coined $\pi$-tons, consist of two particle-hole pairs coupled to the incoming and outgoing light, respectively, and glued together by AFM and CDW fluctuations. Let us emphasize that having two particle-hole pairs (or two holons and two doublons) is a distinct difference to (Mott) excitons \cite{Clarke1993,Wrobel2001,Essler2001,Jeckelmann2003}. 

In other numerical calculations  $\pi$-tons can be identified by doing a channel diagnostics [see SM]. In case of $\pi$-tons it will show the predominance of the particle-hole transversal channel and in case of (Mott) excitons of the particle-hole channel instead. This diagnostics requires only full knowledge of the one- and two-particle Green's functions.

The experimental validation of $\pi$-tons is more challenging.
Indeed, the  optical conductivity has been studied for a  wide range of materials~\cite{BasovRMP2011,Uchida1991, Tajima2016, Lunkenheimer2006, Gossling2008, Gossling2008b,Baldassarre2015, Ruppen2017}. But now that we know that there are $\pi$-tons, too, we need to distinguish these  $\pi$-tons from excitons or in metals from weak localization corrections. We see three routes to do so [see SM for an extended discussion~\footnote{See Supplemental Material [url] for an extended discussion, which includes Refs.~\cite{Pavarini2005,Maier2007,Maier_RPA,Kampf1992,Si2015}}]:\\ 
(1) Employing the characteristics of $\pi$-tons to rely on AFM or CDW fluctuations \footnote{possibly also orbital fluctuations with   ${\mathbf q}=(\pi, \pi, \ldots)$}, we can employ a control parameter such as temperature, uniaxial pressure or a magnetic field to change these fluctuations. If there are $\pi$-tons there will be corresponding changes in the optical spectrum. Indeed such a characteristic change, specifically an unusual reduction of the optical gap around the N\'eel temperature, has been already observed in SmTiO$_3$,\cite{Gossling2008}. To ensure that this effect actually originates from $\pi$-tons excluding a simple reduction of the  one-particle gap  or spin-polaron formations \cite{Martinez1991,Clarke1993,Wrobel2001,Sangiovanni2006,Taranto2012}, additional angular resolved photoemission spectroscopy (ARPES) and inverse ARPES  are necessary as a function of the control parameter. \\
(2) In a joint experimental and theoretical effort we can take the experimental one-particle spectrum $A_{{\bf k}\nu}$ (as e.g.\ measured in  ARPES/inverse ARPES) and the experimental dynamic spin susceptibility $\chi_m({\mathbf q},\omega)$ (as e.g.\  measured by neutron spectroscopy) and calculate from this the optical conductivity $\sigma(\omega)$ including $\pi$-tons and compare it with the measured one. \\
(3) Last but not least we can do {\em ab initio} calculations of strongly correlated materials for which $\pi$-tons may be expected, e.g., along the line of \cite{Galler2016}, and calculate the optical spectrum including $\pi$-tons and exciton contributions. Given good  agreement with experiment and sizable $\pi$-ton effects,
this would provide excellent evidence for  $\pi$-tons.

\acknowledgements {\em Acknowledgements.} We would like to thank Jan Kunes, Gang Li, Angelo Valli and Paul Worm for many stimulating discussions, Josef Kaufmann and Clemens Watzenb\"ock for the help with analytic continuation, and Monika Waas for graphical assistance.
The present work was supported by the 
European Research Council under the 
European Union's Seventh Framework Program
(FP/2007-2013) through ERC Grant No. 306447, the 
Austrian Science Fund (FWF) through  project P 30997-N32 and
Doctoral School ``Building Solids for Function'' (P.P.). Calculations have been done on the Vienna Scientific Cluster (VSC).
\acknowledgements

%


\pagebreak
\clearpage

\onecolumngrid
\begin{center}
  \textbf{Supplemental material: \large Generic optical excitations of correlated systems: $\pi$-tons}\\[.2cm]

{A. Kauch$^{a,*}$, P. Pudleiner$^{a,b,*}$, K. Astleithner$^{a}$, P. Thunstr\"om$^{c}$, T. Ribic$^a$, and K. Held$^a$}\\[.1cm]

{\it $^a$Institute of Solid State Physics, TU Wien, 1040 Vienna, Austria}\\
{\it $^b$Institute of Theoretical and Computational Physics,
Graz University of Technology, 8010 Graz, Austria}\\
{\it $^c$Department of Physics and Astronomy, Materials Theory, Uppsala University, 751 20 Uppsala, Sweden}\\
\thanks{These authors contributed equally to this work}

(Dated: \today)\\[0.3cm]
\end{center}

\setcounter{equation}{0}
\setcounter{figure}{0}
\setcounter{table}{0}
\setcounter{page}{1}
\renewcommand{\theequation}{S\arabic{equation}}
\renewcommand{\thefigure}{S\arabic{figure}}

\begin{center}
\begin{minipage}{.8\textwidth}
In this Supplemental Material, we present in Section A additional results for the optical conductivity  and the corresponding current-current correlation function separated into different  channels  as the parameters of the models are varied. In Section B, we discuss the connection between particle-hole and transversal particle-hole  channel. In Sections C and D, we also show the reducible contributions $\Phi_{\overline{ph}}$  and  $\Phi_{ph}$ to the full vertex $F_{d}$ for the Falicov-Kimball model and the Pariser-Parr-Pople model. In Section E and F we present details on  how to validate $\pi$-tons in numerical data and in experiment, respectively.
\\[0.5cm]
\end{minipage}
\end{center}

\subsection{A. Optical conductivity upon approaching the phase transition}

\begin{figure}[!h]
 \includegraphics[width=0.85\linewidth]{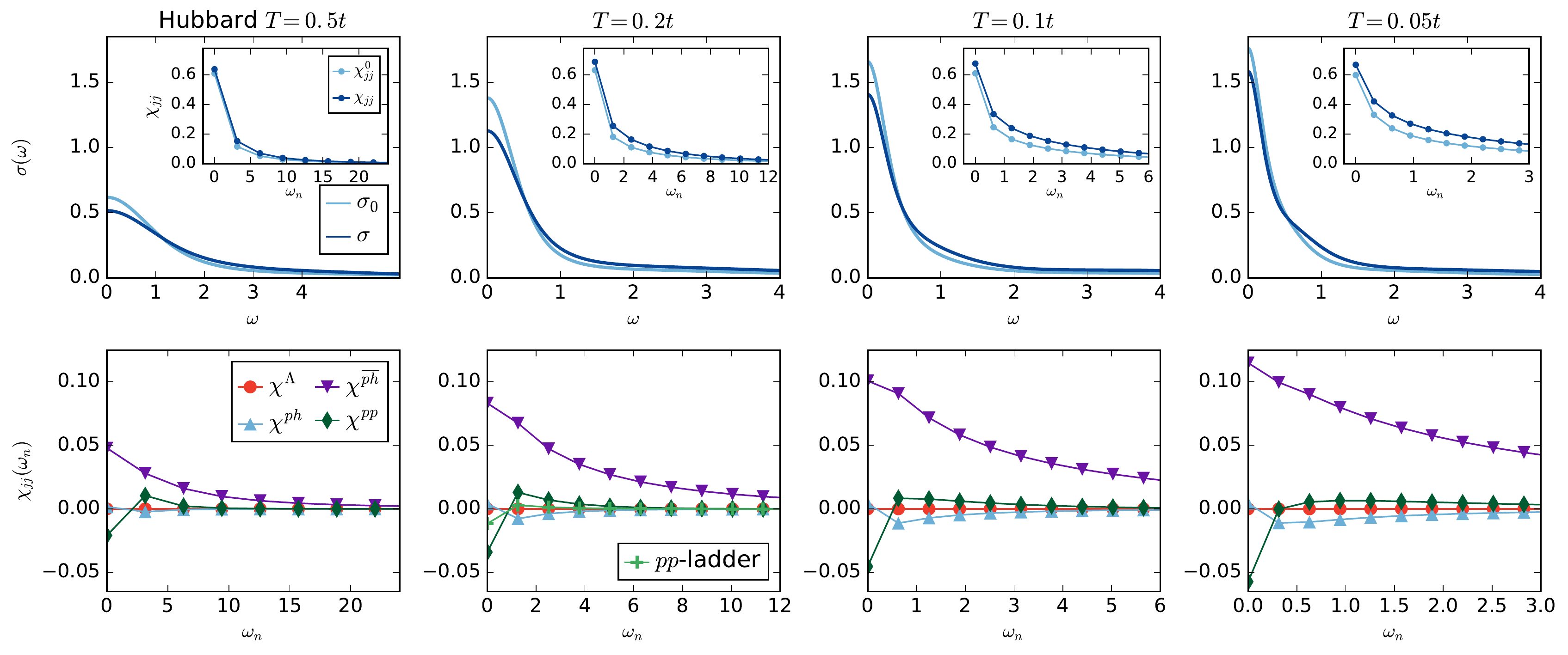}
   \caption{Top: Optical conductivity for real frequency (main panel) and the corresponding current-current correlation function in Matsubara frequencies (insets) of the Hubbard model, showing the bare bubble  ($\sigma_0$) and  the full conductivity ($\sigma$) including vertex corrections (in the insets $\Chi^0_{jj}$ and $\Chi_{jj}$, respectively). Bottom: Corresponding vertex correction to the current-current correlation function $\Chi_{jj}$
separated into the contributions from the three channels  ($ph$, $\overline{ph}$, $pp$) as well as the fully irreducible contribution ($\Lambda$). 
From left to right different temperatures are shown at $U=4t$. For $T=0.2t$  also the contribution of a RPA-like $pp$ ladder is shown.  Other parameters as in the main paper.  \label{fig:optcond_HM}
 }
\end{figure}

\begin{figure}[t]
\vspace{-0.2cm}
 \includegraphics[width=0.715\linewidth]{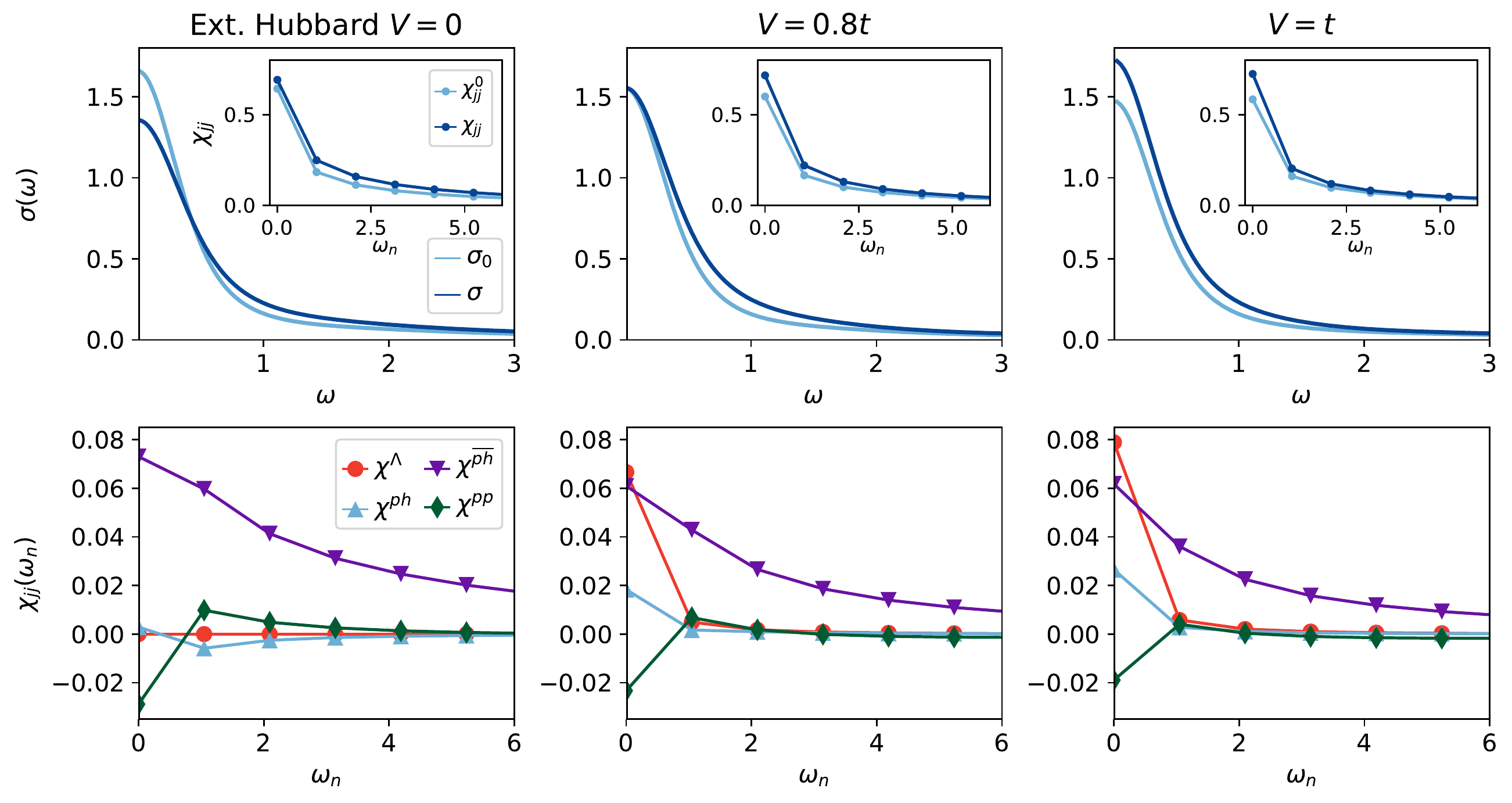}\hspace{0pt}
  \includegraphics[width=0.278\linewidth]{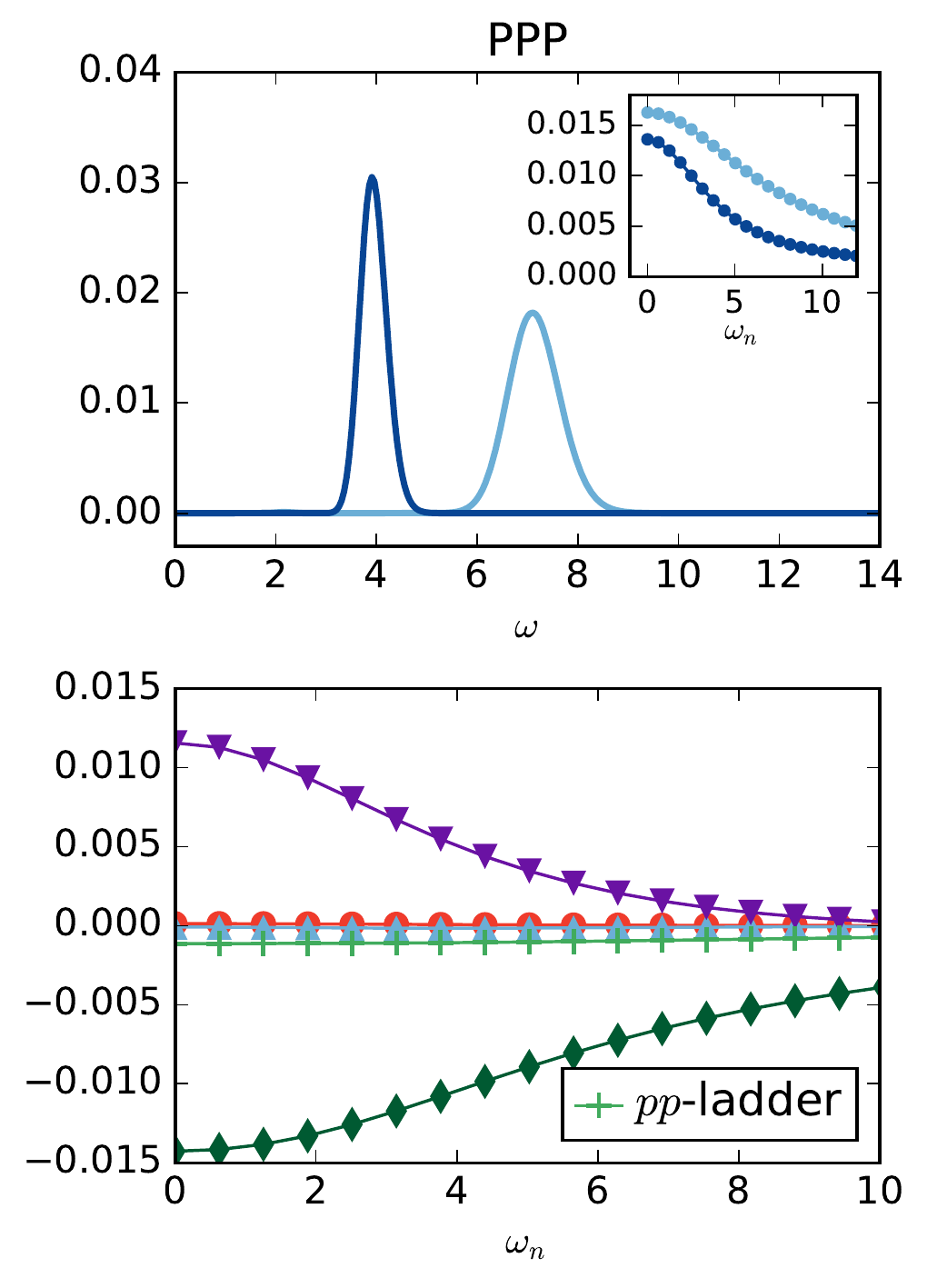}
   \caption{Same as Fig.~\ref{fig:optcond_HM} but for the EHM and PPP model. From left to right different nearest neighbor interactions $V$ are shown. The rightmost column is for the PPP with the electric field along the ring (i.e. $\gamma_{\vek{k}}^{\vek{q}=0}=\partial\epsilon_{\vek{k}}/\partial \vek{k}$). For the PPP also the contribution of a RPA-like $pp$ ladder is shown.  Other parameters are the same as in the main paper.  
 }
   \label{fig:optcond_ExtHM}
\end{figure}

\begin{figure}[!h]
\vspace{-0.2cm}
\begin{minipage}{0.76\linewidth}
\vspace{0pt}
 \includegraphics[width=\linewidth]{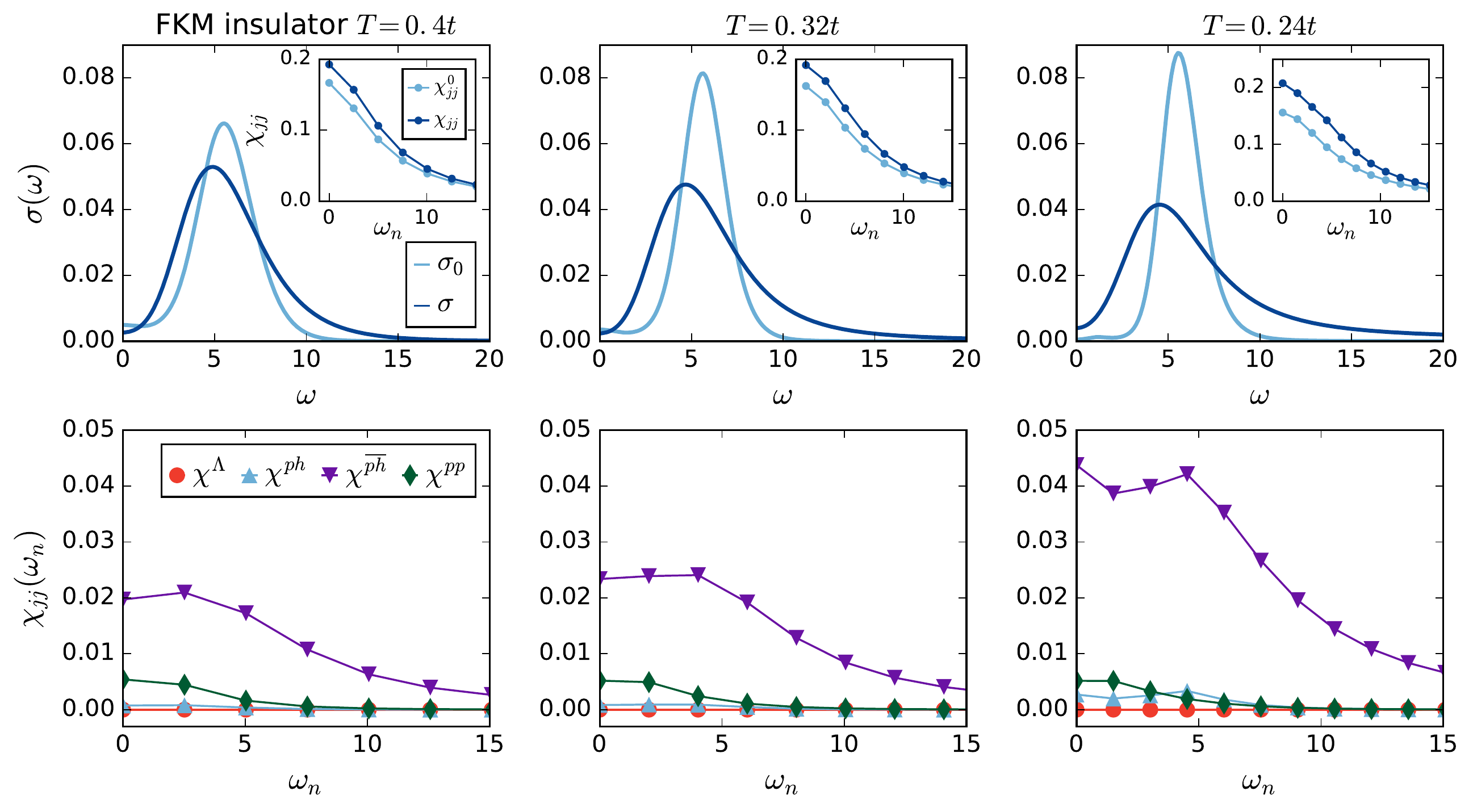}
\end{minipage}\hfill
\begin{minipage}{0.2\linewidth}\vspace{0pt}
   \caption{Same as Fig.~\ref{fig:optcond_HM} but for the FKM at $U=6t$ (insulator). From left to right different temperatures are shown. Other parameters are the same as in the main paper.  
 }
   \label{fig:optcond_FKMins}
\end{minipage}
\end{figure}

\begin{figure}[!h]
\vspace{-0.2cm}
\begin{minipage}{0.76\linewidth}
\vspace{0pt}
 \includegraphics[width=\linewidth]{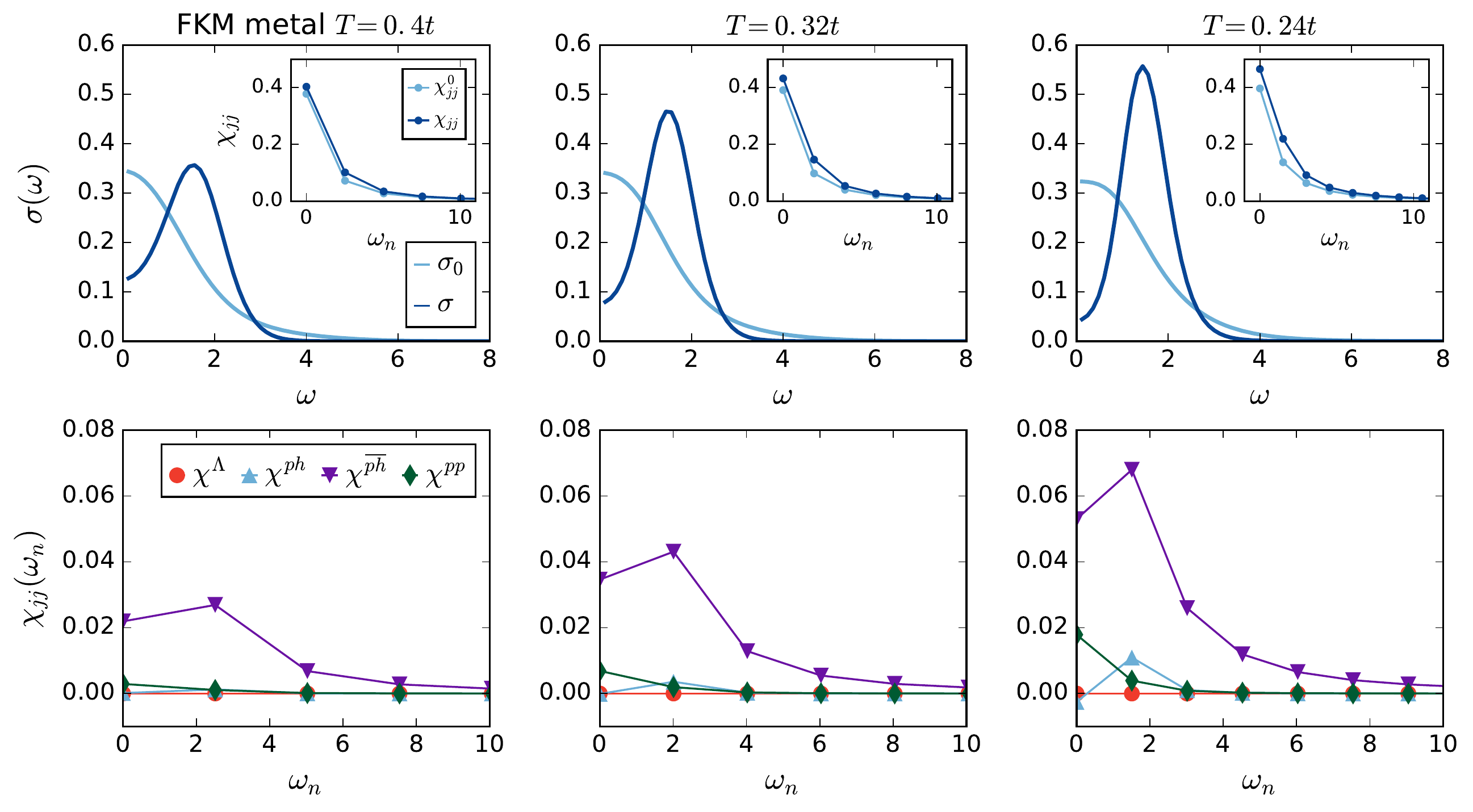}
\end{minipage}\hfill
\begin{minipage}{0.2\linewidth}\vspace{0pt}
   \caption{Same as Fig.~\ref{fig:optcond_HM} but for the FKM at $U=2t$ (metal). From left to right different temperatures are shown. Other parameters are the same as in the main paper.  
 }
     \label{fig:optcond_FKM_metal}
\end{minipage}
\end{figure}

In the Hubbard model (HM; Fig.~\ref{fig:optcond_HM}) and Falicov-Kimball model (FKM; Figs.~\ref{fig:optcond_FKMins}-\ref{fig:optcond_FKM_metal}) we approach the antiferromagnetic (AFM) or charge density wave (CDW) transition by lowering the temperature. In case of the HM the AFM transition is at $T=0$; in case of FKM the CDW transition is around $T=0.12t$ for the metal with $U=2t$ and around $T=0.2t$ for the insulator with $U=6t$~\cite{Ribic2016}.  For the extended HM (EHM; Fig.~\ref{fig:optcond_ExtHM}), on the other hand, we approach the CDW transition by increasing the nearest-neighbor interaction $V$ (the transition to CDW is around $V=1.06t$~\cite{Pudleiner2019a}).

In all cases studied, as we approach the phase transition the vertex contribution becomes larger. As shown in the bottom panels of Figs.~\ref{fig:optcond_HM}-\ref{fig:optcond_FKM_metal} the largest overall contribution to the vertex part is $\Chi^{\overline{ph}}$ [except for the Pariser-Parr-Pople (PPP) model  in Fig.~\ref{fig:optcond_ExtHM}, where it is   $\Chi^{pp}$], with the second biggest being  $\Chi^{pp}$ (with the exception of the EHM in Fig.~\ref{fig:optcond_ExtHM}, where it is   $\Chi^{\Lambda}$). However, as explained in the main text, the large values of $\Chi^{pp}$ mainly come from insertions of other diagrams into the pure RPA-like $pp$-ladder that are generated in the parquet solution (for comparison see the bottom middle left plot in Fig.~\ref{fig:optcond_HM} and the bottom right of Fig.~\ref{fig:optcond_ExtHM}, where the contribution from a $pp$-ladder without other diagrammatic insertions is shown).




\subsection{B. Relation between the $ph$ and $\overline{ph}$ reducible contributions to the full vertex $F_d$}

As we have shown in the main paper and in the previous Section, the transversal particle-hole ($\overline{ph}$) channel contains the most important vertex corrections to the optical conductivity.  On the other hand this $\overline{ph}$ channel has just  the same structure as the  $ph$ channel. Indeed, it is the rotated version of that channel. The optical conductivity contribution is only different because, when connected with the electron-photon interaction lines, we select different momenta of the two respective channels.

Specifically, the  $\overline{ph}$-reducible vertex at ${\mathbf q}=0$ is connected to the  ${ph}$-reducible vertex  vertices as follows
\begin{equation}
\Phi_{\overline{ph},d,{\vek{k}\vek{k'}\vek{q}=0}}^{\nu_n\nu_n'\omega_n}=-\frac{1}{2}(\Phi_{{ph},d}+3\Phi_{{ph},m})_{{\vek{k}\,\vek{k}\,{\vek{k'}-\vek{k}}}}^{\nu_n\nu_n+\omega_n\nu_n'-\nu_n}
\label{eq:reducible}
\end{equation}
 with $d$ ($m$) denoting the even (odd) spin combination, see \cite{RMPVertex,Li2016} for more details.
This above combination of arguments of the  ${ph}$-reducible vertex $\Phi_{{ph},d/m}$ is entering the optical conductivity as the transversal particle-hole channel, whereas $\Phi_{ph,d,{\vek{k}\vek{k'}\vek{q}=0}}^{\nu_n\nu_n'\omega_n}$ is the contribution that enters as the particle-hole channel when calculating the optical conductivity.


\subsection{C. Reducible contributions to the full vertex $F_d$ in the $ph$ and $\overline{ph}$ channels for the FKM} 

\begin{figure} [!h]
\vspace{-0.2cm}
\begin{minipage}{0.72\linewidth}
\vspace{0pt}
 \includegraphics[width=\linewidth]{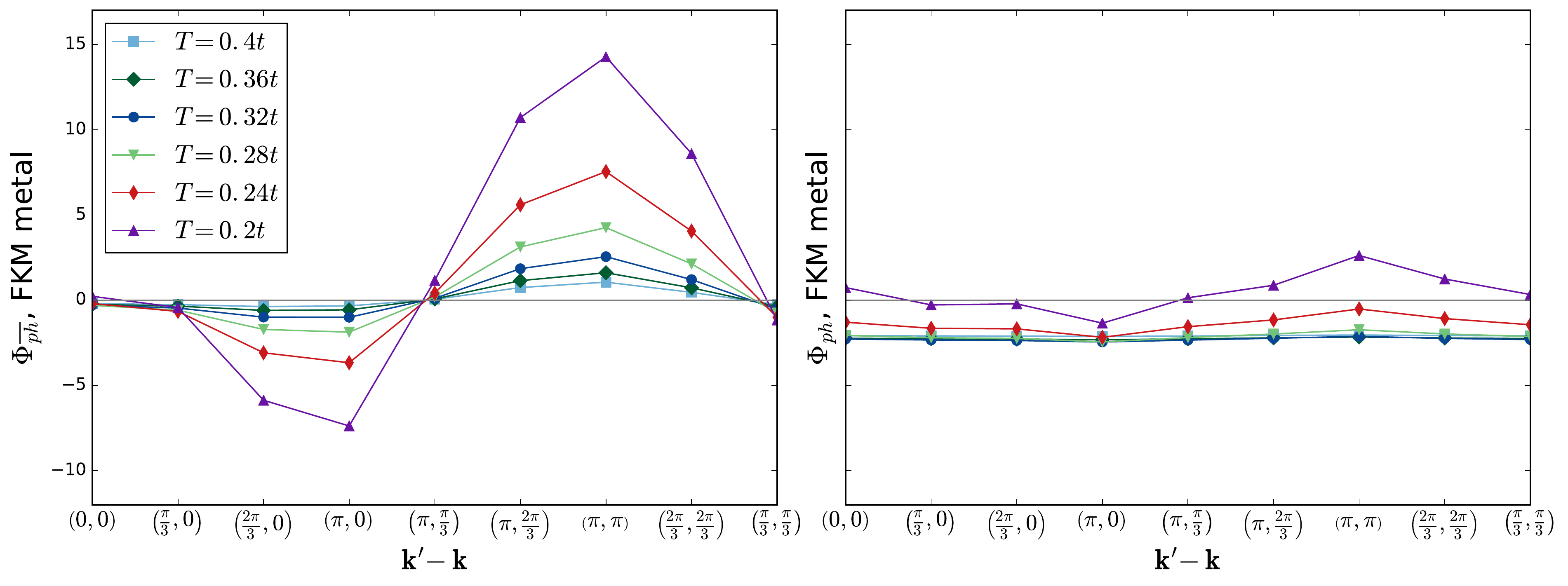}
\end{minipage}\hfill
\begin{minipage}{0.24\linewidth}\vspace{0pt}
   \caption{Reducible contributions $\Phi$  in the $\overline{ph}$ (left) and $ph$ (right) channel to the full vertex $F_{d,{\vek{k}\vek{k}'\vek{q}=0}}^{\nu_n\nu_n'\omega_n}$ correcting the optical conductivity for the FKM at various temperatures and $U=2t$ (metal).  Shown is the contribution $\nu_n= \nu_n'= \pi T$; $\omega_n=0$ at fixed $\vek{k}=(0,0)$ as a function of $\vek{k}'-\vek{k}$.
   \label{fig:phi_FKM}}
\end{minipage}
\end{figure}

\begin{figure} [!h]
\vspace{-0.2cm}
\begin{minipage}{0.72\linewidth}
\vspace{0pt}
 \includegraphics[width=\linewidth]{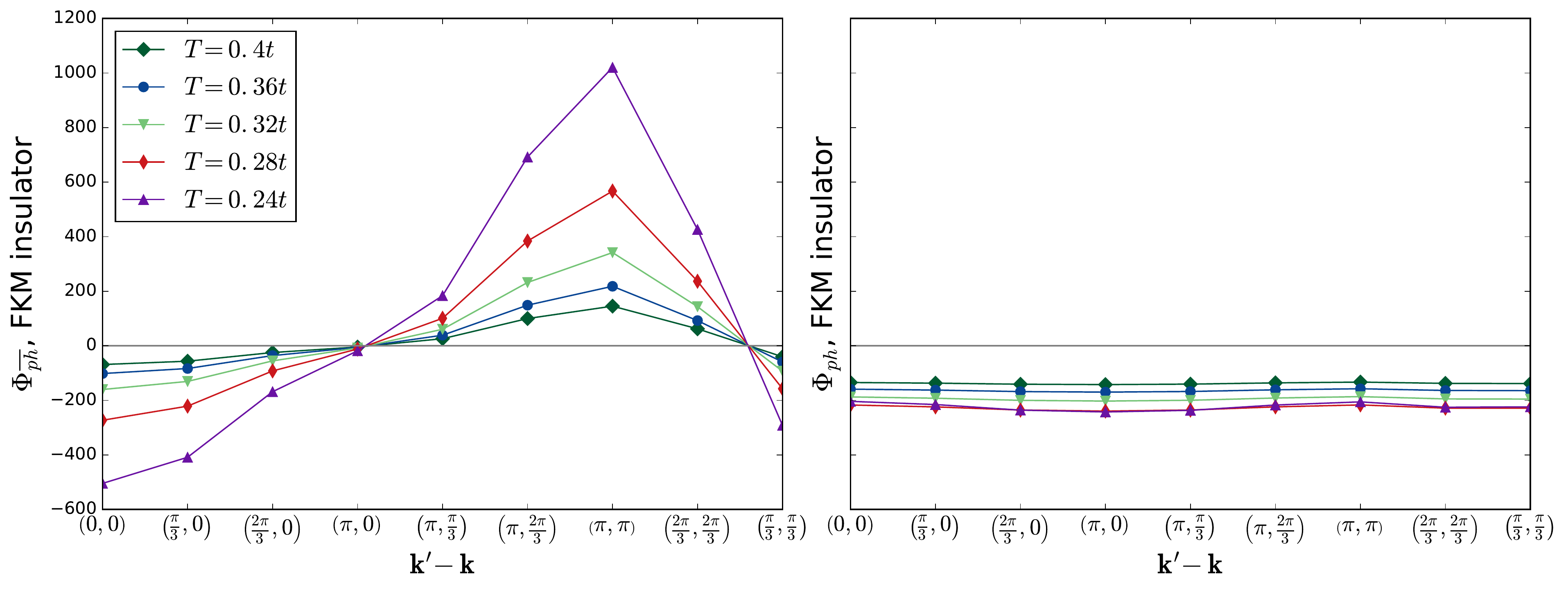}
\end{minipage}\hfill
\begin{minipage}{0.24\linewidth}\vspace{0pt}
   \caption{The same as Fig.~\ref{fig:phi_FKM} but for $U=6t$ (insulator).  
   \label{fig:phi_FKMi}}
\end{minipage}
\end{figure}

We also show in Figs.~\ref{fig:phi_FKM}-\ref{fig:phi_FKMi} the reducible contributions $\Phi_{\overline{ph}}$  and  $\Phi_{ph}$ to the full vertex $F_{d}$ for the FKM in the metallic (Fig.~\ref{fig:phi_FKM}) and insulating (Fig.~\ref{fig:phi_FKMi}) phase. Also in this case, similarly as for the HM, EHM and PPP model shown in the main text, the $\vek{k}'-\vek{k}=(\pi,\pi)$ contribution to $\Phi_{\overline{ph}}$ is the largest. As we approach the phase transition, it grows, whereas  $\Phi_{ph}$ stays small and does not depend on  $\vek{k}'-\vek{k}=(\pi,\pi)$ much, except for the lowest temperature of $T=0.2t$ in Fig.~\ref{fig:phi_FKM} where a small  $\vek{k}'-\vek{k}$ dependence develops.

 Let us note that the FKM is difficult to solve in D$\Gamma$A as the latter requires a mixed vertex with mobile and immobile electrons \cite{Ribic2016,Ribic2016b}. This is the reason why we employ the dual fermion approach for the FKM.

\subsection{D. Reducible contributions to the full vertex $F_d$ in the $ph$ and $\overline{ph}$ channels for the PPP model} 
In the main paper, we have shown that the reducible vertex contributions for the HM and extended model are strongly  peaked at  $\vek{k}'-\vek{k}=(\pi,\pi)$, and in the previous section that the same holds for the FKM.
Here, we turn to the PPP model, which has a one-dimensional momentum vector since it can be considered as a Hubbard model on a six site ring with additional non-local interactions to all sites. This allows us to show the dependence of both one-dimensional vectors,  $\vek{k}'-\vek{k}=(\pi,\pi)$ and $\vek{k}$ (or, alternatively,  $\vek{k}'$ and $\vek{k}$), with  six ${\mathbf k}$-points each, as well as on  the Matsubara frequency in a three-dimensional plot.

\begin{figure}
\vspace{-0.2cm}
\vspace{0pt}
\begin{center}
 \includegraphics[width=0.8\linewidth]{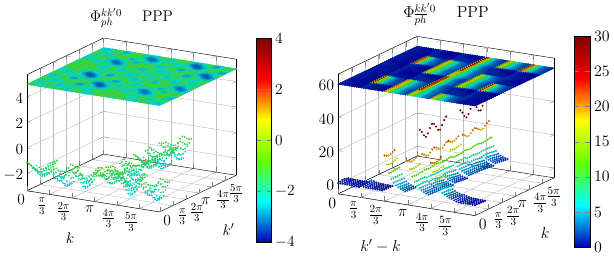}
\end{center}
   \caption{(Color online) Reducible contributions $\Phi$  in the $ph$ (left) and $\overline{ph}$-channel  (right) entering the optical conductivity  as vertex corrections plotted vs.~the four-vectors  $k$ and ${k}'$ or ${k}'-{k}$, respectively. For each combination of momenta, the  dependence on frequencies is plotted on the finer scale (bosonic frequency $\omega_n=0$). Parameters as in Fig.~\ref{fig:optcond} of the main paper, i.e., 
$T=0.1t$, $U =3.962t$, $V_{01}=2.832t$,   $V_{02}= 2.014t$,  $V_{03} = 1.803t$.
 \label{Fig:phi_3d}}  
\end{figure}

This plot is shown in Fig.~\ref{Fig:phi_3d}. Again we see that the  $\overline{ph}$-channel is largest with a pronounced peak at  $\vek{k}'-\vek{k}=(\pi,\pi)$. This confirms the picture of the $\pi$-tons feeding upon strong AFM or CDW fluctuations.

Let us also note that, despite the size of its gap, the PPP model is not a Mott insulator but has a mildly strong  self-energy \cite{Pudleiner2019}.

\subsection{E. Identification of $\pi$-tons in other numerical methods}
In this Section, we would like to briefly outline how to analyze numerical data for the existence of $\pi$-tons. A first indication and hint is the ionicity (number of doublons relative to the ground state) of the optically excited states, specifically those affected by vertex corrections (e.g.\ states below the one-particle gap).  If a wave function is given for this excited state as in exact diagonalization (ED), density matrix renormalization group (DMRG) etc. the number of doublons can be readily calculated from this wave function. 

\begin{figure}[!h]

\vspace{-0.2cm}
\begin{minipage}{0.4\linewidth}
\vspace{0pt}
 \includegraphics[width=\linewidth,trim=0cm 6cm 0cm 0cm, clip ]{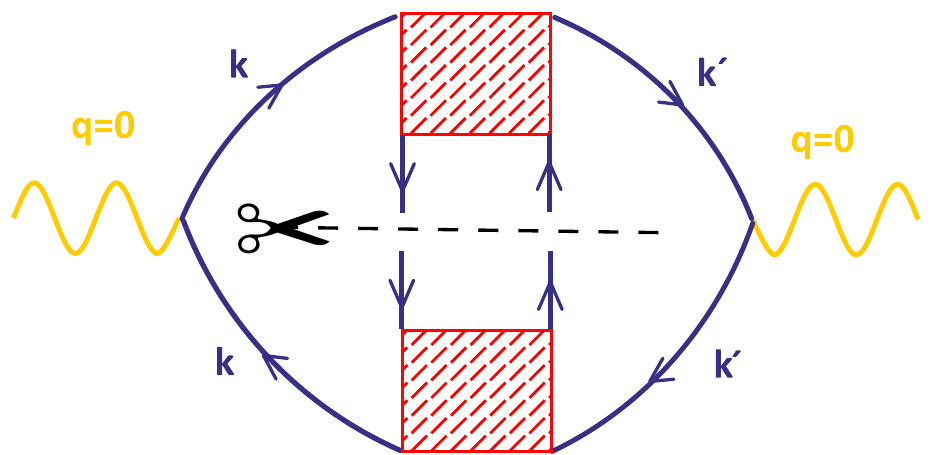}
\end{minipage}\hfill
\begin{minipage}{0.58\linewidth}\vspace{0pt}
   \caption{$\pi$-tons necessitate two particles and two holes, because this is the only way to construct a diagram in the $\overline{ph}$-reducible channel connected to incoming and outgoing light (yellow wiggly lines). In any diagram of the reducible $\overline{ph}$ channel we can cut two Green's function lines (blue lines) as indicated, and have Coulomb interactions within the dashed red vertex boxes. Since (i) we still need one particle and one hole to finally connect to the outgoing light, and since (ii) both of them still need at least one interaction in the red boxes, and (iii) cannot interact with each other any more because of the cut lines, we have at least two particles and two holes for the $\pi$-ton.      
 }
   \label{fig:piton_diagram}
\end{minipage}
\end{figure}

If  we are well in the  Mott insulating state so that doublons are otherwise suppressed, this ionicity should  be approximately one doublon in case of (Mott) excitons  \cite{Essler2001,Jeckelmann2003}  including spin Peierls physics \cite{Gossling2008,Wrobel2001,Clarke1993}  and two doublons or more if $\pi$-tons are present (see Fig.~\ref{fig:piton_diagram})~\footnote{Other processes with two doublons considered so far are Aslamazov-Larkin diagrams where also two interaction lines can be substituted by the antiferromagnetic spin susceptibility~\cite{Kampf1992}. Such a diagram is part of the particle-hole transversal channel but if we couple the antiferromagentic spin susceptibility this way, there is no strong peak around ${\mathbf k}-{\mathbf k'}=\pi$ in contrast to Fig.~3. Also biexcitons, which include two particles and two hoels as well, do not show such a ${\mathbf k}-{\mathbf k'}$-dependence.}. For this reason  Mott excitons should also be more important than  $\pi$-tons around the atomic limit. But when there are large vertex corrections  in the one-dimensinal EHM, i.e.\ for $2V\lesssim U$, the ionicity in DMRG is too large for plain Mott excitons \cite{Jeckelmann2003}.

A further difference to (Mott) excitons and spin Peierls physics is that in these pictures electron and hole have a total momentum ${\bf q}=0$. This might help avoiding breaking up AFM bonds in the spin Peierls picture. Wheras for teh $\pi$-ton the electron and hole at the cut line in Fig.~\ref{fig:piton_diagram} may carry a total momentum of  ${\bf q}=\pi$. This way the spin background can be rotated just as in an AFM paramagnon.

More direct evidence can be obtained by doing a diagnostics of the contribution of the different channels. To this end, the one-particle Green's function $G_{{\bf k}\nu}$  and two-particle Green's function $G_{\sigma\sigma',\mathbf{k}\mathbf{k'}\mathbf{q}}^{(2),\nu\nu'\omega}$ are needed as input from any numerical calculation, see \cite{RMPVertex} for the frequency definition and further details. In this Section we denote the Matsubara frequencies by $\nu$ and $\omega$, i.e. we skip the subscript $n$. 

(1) From $G$ and $G^{(2)}$,  we can calculate the generalized  susceptibilities  in the paramagnetic phase for the density ($d$) and magnetic ($m$) component as follows:
\begin{eqnarray} 
 \label{equ:defgensusc} 
 \Chi_{\sigma\sigma',\mathbf{k}\mathbf{k'}\mathbf{q}}^{\nu\nu'\omega}&=&G_{\sigma\sigma',\mathbf{k}\mathbf{k'}\mathbf{q}}^{(2),\nu\nu'\omega}-\beta G_{\mathbf{k}\nu}G_{\mathbf{k'}\nu'}\delta_{\omega 0}\delta_{\mathbf{q}\mathbf{0}}\nonumber\\
 \Chi_{d/m,\mathbf{k}\mathbf{k'}\mathbf{q}}^{\nu\nu'\omega}&=&\Chi_{\uparrow\uparrow,\mathbf{k}\mathbf{k'}\mathbf{q}}^{\nu\nu'\omega}\pm\Chi_{\uparrow\downarrow,\mathbf{k}\mathbf{k'}\mathbf{q}}^{\nu\nu'\omega}. 
\end{eqnarray}

(2) These susceptibilities are connected to the full vertex $F$ via
\begin{eqnarray} 
 \label{equ:defvertex} 
 \Chi_{d/m,\mathbf{k}\mathbf{k'}\mathbf{q}}^{\nu\nu'\omega}&=&{{-\beta G_{\mathbf{k}\nu}G_{(\mathbf{k}+\mathbf{q})(\nu+\omega)}\delta_{\nu\nu'}\delta_{\mathbf{k}\mathbf{k'}}}}- G_{\mathbf{k}\nu} G_{(\mathbf{k}+\mathbf{q})(\nu+\omega)}F_{d/m,\mathbf{k}\mathbf{k'}\mathbf{q}}^{\nu\nu'\omega}G_{\mathbf{k'}\nu'}G_{(\mathbf{k'}+\mathbf{q})(\nu'+\omega)},
\end{eqnarray}
and we can resolve this equation for $F$. Here, the division by  $G$'s will lead to quite some noise at large (Matsubara) frequency. But the dependence of the final result, the channel-decomposed optical conductivity on this noise is rather mild.

(3) Having $F$ and the one-particle Green's function, we can (matrix) invert the Bethe-Salpeter equation in all three channels to obtain the particle-hole irreducible vertex $\Gamma$ in each channel.  Let us state the Bethe-Salpeter equation only for the $ph$-channel explicitly 
\begin{eqnarray} 
 \label{equ:decomp2PIGamma} 
  F_{d/m,\mathbf{k}\mathbf{k'}\mathbf{q}}^{\nu\nu'\omega}&=&\Gamma_{ph,d/m,\mathbf{k}\mathbf{k'}\mathbf{q}}^{\nu\nu'\omega}+\frac{1}{\beta N}\sum_{\mathbf{k_1}\nu_1}\Gamma_{ph,d/m,\mathbf{k}\mathbf{k_1}\mathbf{q}}^{\nu\nu_1\omega} G_{\mathbf{k_1}\nu_1}G_{(\mathbf{k_1}+\mathbf{q})(\nu_1+\omega)}F_{d/m,\mathbf{k_1}\mathbf{k'}\mathbf{q}}^{\nu_1\nu'\omega}.\nonumber
\end{eqnarray}

(4) From $F$ and  $\Gamma_\ell$, we obtain the reducible vertex  $\Phi_{\ell, d/m,\mathbf{k}\mathbf{k'}\mathbf{q}}^{\nu\nu'\omega}=F_{d/m,\mathbf{k}\mathbf{k'}\mathbf{q}}^{\nu\nu'\omega}-\Gamma_{\ell, d/m,\mathbf{k}\mathbf{k'}\mathbf{q}}^{\nu\nu'\omega}$ in each channel $\ell \in\{ph, \overline{ph}, pp\}$.

(5) Now we set  $F=\Phi_\ell$ and calculate with Eq.~(2) of the main text the contribution of the $\ell=ph$ channel (excitons), $\ell=\overline{ph}$ channel ($\pi$-tons) and $\ell=pp$ (cooperons/weak localization). This way we can detect which kind of physics is dominant. A further analysis which momenta yield the main contributions, as in our paper, is possible as well.

\subsection{F. Experimental validation of $\pi$-tons}

Let us start by mentioning that a strongly correlated transition metal oxide,
 SmTiO$_3$,\cite{Gossling2008} might actually show $\pi$-tons in its optical spectrum. More precisely,  SmTiO$_3$ shows a reduction of the optical gap like in our calculations --- around the $T$-induced antiferromagnetic transition. 
Since the one-particle gap is expected to increase upon 
antiferromagnetic order, and many other mechanism could be 
ruled out \footnote{G{\"o}ssling {\em et al.} \cite{Gossling2008} show that the first peak of the optical conductivity is within the one-particle gap, so that the authors ruled out Mott excitons as discussed in \cite{Essler2001,Jeckelmann2003} since  unrealistically large (for a transition metal oxide) 
nearest-neighbor interaction are required for such an 
in-gap optical peak within the Mott exciton picture. 
G{\"o}ssling {\em et al.} \cite{Gossling2008}  
instead consider a holon and doublon coupled in a 
spin-Peierls-like mechanism \cite{Wrobel2001,Clarke1993}. 
But if this variant of a Mott exciton is present it should 
foremost result in the development of additional 
characteristic (T-dependent) peaks in the optical spectrum not a shift of the gap edge \cite{Taranto2012}, exactly as it was observed in LaSrMnO$_4$ \cite{Gossling2008b}.},
this hints towards $\pi$-tons (which reduce the gap and 
rely on strong antiferromagnetic fluctuations). However, this is certainly 
not a full fledged proof for $\pi$-tons since as (quite typical 
in solid state physics) also all kind of other effects such as 
e.g.\ an orbital disproportionation that affects the one-particle gap might be present.

A smoking gun experiment is difficult to pursue but nonetheless doable.

(1) One possibility is to employ our knowledge that the $\pi$-tons rely on antiferromagnetic and charge density wave fluctuations (in contrast to all other known polaritons such as excitons and Mott excitons), and to change the strength of these fluctuations by an external parameter. Essentially that is what the aforementioned experiments for SmTiO$_3$ did with the parameter being temperature. Other possible parameters are e.g. uniaxial pressure or, for antiferromagnetic fluctuations, a magnetic field. But to be sure that the effect is not a one-particle effect, one needs (on top of what one has done for SmTiO$3$) to perform angular resolved photoemission spectroscopy (ARPES)/inverse ARPES to be sure that the observed reduction of the optical gap is not a one-particle effect. 

In the case of SmTiO$_3$, such ARPES experiments will also confirm (or disconfirm)  the spin-polaron Mott exciton picture of  \cite{Wrobel2001,Clarke1993,Gossling2008}. This picture has characteristic peaks in the optical conductivity {\em and} in the one-particle spectrum \cite{Martinez1991,Sangiovanni2006}. In the optical conductivity of SmTiO$_3$ such peaks are not visible but it is certainly most valuable to confirm this by ARPES. While we think that   SmTiO$_3$ is a good candidate, as a matter of course other strongly correlated materials can be investigated along this first route to validate $\pi$-tons experimentally.

As for the control parameters, please note, that in case of a magnetic field one can, as a matter of course, not do ARPES; and the energy scale of typical magnetic  field strengths is only sufficient to suppress antiferromagnetism with a N\'eel temperature of a few Kelvin. Hence temperature or uniaxial pressure appear to be more suitable control parameters. Note that by uniaxial pressure one can  e.g. split  the $t_{2g}$ energy levels in a $d^ 1$ configuration. Without splitting one $d$ electron in three $t_{2g}$ supports  a paramagnetic or ferromagnetic configuration. With splitting, one has a half-filled situation with one $d$ electron in one (split-off) $t_{2g}$ level which is favorable for antiferromagnetism. \footnote{A similar effect to uniaxial pressure occurs in a dimensional reduction through heterostructures. It leads e.g. in SrRuO$_3$ to a change from a ferromagnetic to an antiferromagnetic phase  \cite{Si2015}}

(2) A second possibility is to do a third experiment (besides ARPES and optics) measuring the antiferromagnetic spin susceptibility e.g. with neutron spectroscopy. From the ARPES bandstructure and the spin susceptibility we can calculate the optical spectrum. If this agrees with experiment and the optical gap sensitively depends on the $\pi$-ton contribution, we have an elegant proof of $\pi$-tons based on a joint theoretical and experimental study.

This validation is based on our observation that the by far most important contribution to the vertex corrections in the optical conductivity, the $\pi$-ton, stems from the transversal particle-hole channel. For antiferromagnetic fluctuations this can, in turn, be calculated to a good accuracy from the  dynamic antiferromagnetic spin susceptibility.

 Since neutron spectroscopy is restricted to low energies (meV to hundreds meV's), such an experiment is best done for an antiferromagnet with a low N\'eel temperature so that one does not miss important contributions in the dynamic spin susceptibility.

 Let us briefly sketch the calculation needed for this validation: From the (inverse) ARPES, we directly get the spectrum $A_{{\mathbf k} \nu}$ and from this  the Green's function $G_{{\mathbf k}\nu}$. From the experimental susceptibility $\Chi_m({\mathbf q},\omega)$ in turn we get the estimate of the magnetic $ph$-reducible vertex $\Phi_{ph,m}$. Under the assumption that the AFM fluctuations are strong, the bosonic momentum and frequency dependence will dominate giving~\footnote{Here we assume $\Chi_m\gg\Chi_0$ and neglect the terms with the bare susceptibility  $\Chi_0=-\frac{1}{\beta N}\sum_{\nu''\,{\bf k}''} G_{\mathbf{k}''\nu''}G_{(\mathbf{k}''+\mathbf{q})(\nu''+\omega)}$, but they can also be explicitely included.}
\begin{equation}
\Phi_{{ph},m,{\vek{k}\,\vek{k'}\,\vek{q}}}^{\nu\nu'\omega}\cong \Phi_{{ph},m,}({\bf q},\omega) =
-\frac{\Chi_m({\mathbf q},\omega)}{\left(\frac{1}{\beta N}\sum_{\nu''\,{\bf k}''} G_{\mathbf{k}''\nu''}G_{(\mathbf{k}''+\mathbf{q})(\nu''+\omega)} \right)^2}.
\end{equation}
Asuming the dominance of AFM fluctuations over density fluctuations, we can further write, after Eq.~\ref{eq:reducible},
\begin{equation}
\Phi_{\overline{ph},d,{\vek{k}\,\vek{k'}\,\vek{q}=0}}^{\nu\nu'\omega}\cong 
-\frac{3}{2} \Phi_{{ph},m,}({\bf k} - {\bf k}',\nu-\nu').  
\end{equation} 
To obtain only the $\pi$-ton contribution  to the optical conductivity we finally need to insert
 \begin{equation}
F_{d,{\vek{k}\vek{k'}\vek{q}=0}}^{\nu\nu'\omega}=\Phi_{\overline{ph},d,{\vek{k}\vek{k'}\vek{q}=0}}^{\nu\nu'\omega}
\end{equation}
 and the Green's function $G_{{\mathbf k}\nu}$ into Eq.~(2)~\footnote{One can either translate the Green's function and susceptibility from the experimental data into Matsubara frequencies or use a real frequency version of Eq.~(2)} in the main text to calculate the current-current correlation function and from this the optical conductivity. 

Let us note that a similar approach is followed for high-temperature superconductivity, where the Eliashberg equation is solved with the interaction vertex calculated from magnetic susceptibility as an input~\cite{Maier2007,Maier_RPA}.

(3) A third possibility is to do {\em ab initio} calculations with predictive power for a material with an optical spectrum that strongly depends on a parameter such as temperature, and to compare the calculated optical spectrum (including $\pi$-tons \footnote{This is possible e.g. by advanced numerical methods such as the {\em ab initio} dynamical vertex approximation  \cite{Galler2016}.}) and with experiment. If the optical spectrum  {\em and} the  parameter-dependence agree well, this is compelling proof that the effect originates from $\pi$-tons. This would also be a possibility for SmTiO$_3$, for which density functional theory plus dynamical mean-field theory (DFT+DMFT) calculations without vertex corrections \cite{Pavarini2005,Gossling2008} show a way too large optical gap.  Do vertex corrections, do $\pi$-tons reduce this to the experimentally measured gap? Such a calculation is feasible but a major endeavor, and hence beyond the scope of the present paper.

\end{document}